\documentclass[%
reprint,
superscriptaddress,
amsmath,amssymb,
aps,
pre,
colorlinks
]{revtex4-2}

\usepackage{graphicx}
\usepackage{dcolumn}
\usepackage{bm}

\usepackage[usenames,dvipsnames]{xcolor}
\usepackage[urlcolor=NavyBlue,citecolor=NavyBlue,linkcolor=NavyBlue]{hyperref}

\usepackage[disable]{todonotes}
\usepackage{physics}
\usepackage{bm}
\usepackage{upgreek}

\newcommand{\defeq}{\mathrel{\mathop:}=}
\newcommand{\eqdef}{=\mathrel{\mathop:}}
\newcommand{\ddint}[1]{\!\dd{#1}}
\renewcommand{\vec}[1]{\boldsymbol{ #1 }}

\newcommand{\changed}[1]{{\leavevmode\color{black}{#1}}}

\graphicspath{{./figs/}}

\begin{document}

	\preprint{APS/123-QED}
	
	\title{Likelihood-Based Heterogeneity Inference Reveals Non-Stationary Effects in Biohybrid Cell-Cargo Transport}
	
	\author{Jan Albrecht}
	\email{jan.albrecht@uni-potsdam.de}
	\affiliation{Institute of Physics and Astronomy, University of Potsdam, 14476 Potsdam, Germany}
    
	\author{Lara S. Dautzenberg}
	\affiliation{Institute of Physics and Astronomy, University of Potsdam, 14476 Potsdam, Germany}

    \author{Manfred Opper}
	\affiliation{Faculty of Electrical Engineering and Computer Science, Technische Universit\"at Berlin, 10587 Berlin, Germany}
 	\affiliation{Institute of Mathematics, University of Potsdam, 14476 Potsdam, Germany}
	\affiliation{Centre for Systems Modelling and Quantitative Biomedicine, University of Birmingham, Birmingham B15 2TT, United Kingdom}%

    \author{Carsten Beta}%
	\email{beta@uni-potsdam.de}
	\affiliation{Institute of Physics and Astronomy, University of Potsdam, 14476 Potsdam, Germany}
	\affiliation{Nano Life Science Institute (WPI-NanoLSI), Kanazawa University, Kakuma-machi, Kanazawa 920-1192, Japan}
    
	\author{Robert Gro{\ss}mann}%
	\email{rgrossmann@uni-potsdam.de}
	\affiliation{Institute of Physics and Astronomy, University of Potsdam, 14476 Potsdam, Germany}

	\date{\today}

	\begin{abstract}
	Variability of motility behavior in populations of microbiological agents is a ubiquitous phenomenon even in the case of genetically identical cells. 
	Accordingly, passive objects introduced into such biological systems and driven by them will also exhibit heterogeneous motion patterns.
	Here, we study a biohybrid system of passive beads driven by active ameboid cells and use a likelihood approach to estimate the heterogeneity of the bead dynamics from their discretely sampled trajectories. We showcase how this approach can deal with information-scarce situations and provides natural uncertainty bounds for heterogeneity estimates. Using these advantages we particularly uncover that the heterogeneity in the system is time-dependent.
\end{abstract}
		
	\maketitle
	
	\section{Introduction}
%
%
	
During major concerts or other festivities with densely packed crowds, organizers occasionally distribute large balloons for the attendees to bounce around and interact with~\cite{Gu2025, Orbital2024}. 
This is an example of a macroscopic active system interacting with and collectively driving passive objects; this paper studies a very similar system---albeit shrunk by 5 orders of magnitude. 

Systems of self-propelled motile individuals~\cite{Romanczuk2012} on their own already exhibit a wide range of interesting collective phenomena~\cite{te2025metareview} like large-scale patterns~\cite{Marchetti2013} and non-equilibrium phase transitions~\cite{Cates2015,chate_dadam_2020,baer_self_2020,alert_active_2022}. 
Since real-life active systems are rarely clean and isolated, the interaction of such active particles with obstacles and movable passive objects is an active area of research~\cite{Bechinger2016}. In this regard, much work has focused on the diffusion of passive tracer particles in suspensions of micro-swimmers~\cite{Shea2024, Leptos2009, GuzmanLastra2021}. 
Another important research direction is studying composite systems, in which there is direct physical contact between the active and passive particles rather than hydrodynamic interaction~\cite{CarlsenSitti2014, Webster-Wood2023, Weibel2005,Grossmann2024, altshuler_environmental_2024}. While there are interesting effects in macroscopic realizations of such composite systems, too, the research focus has been on microscopic systems, where they have potential applications in micro-manipulation and -transport~\cite{Sokolov2010, diLeonardo_bacterial_2010, Martel2010} as well as in targeted drug delivery~\cite{Nagel2019, Park2013}. To exploit these composite systems, a thorough understanding and modeling of the motion patterns of their constituents is required.

A ubiquitous aspect of biological active matter is its inherent inter-individual variability that is present even for genetically identical microorganisms~\cite{Meacham2013, Peled2021, Ariel2022} and can manifest itself in their motility~\cite{Spudich1976, Waite2016, Klimek2025}. When passive objects are driven by a heterogeneous population of cells, the objects' dynamics inherit the heterogeneity, which may be further increased by fluctuations in the number of cells attached to each object~\cite{Grossmann2024}. Such population heterogeneity can pose challenges for model inference, as it complicates the interpretation of population-averaged quantities and can lead to unusual statistics, such as non-Gaussian displacement distributions~\cite{Lemaitre2023, Grossmann2024, Cherstvy2018}. Therefore, inferring the variability of a system constitutes an important part of its characterization. 

Likelihood-based approaches can be used to obtain heterogeneity estimates directly from the measured trajectory data, rather than via an intermediate estimation of motility parameters~\cite{Albrecht2024, DelattreLavielle2013, Picchini2010}. Estimators derived from the likelihood, such as maximum-likelihood estimators (MLEs), have favorable statistical properties and provide natural uncertainty estimates for the inferred parameters~\cite{Schervish1995, DelattreLavielle2013}. Furthermore, these methods perform particularly well in situations with little available data, which is why they are also a popular tool in pharmacological research where they are used to infer inter-patient variability~\cite{Donnet2013}. 

In this article, we consider an experimental biohybrid cell-cargo system consisting of a carpet of \textit{Dictyostelium discoideum}~(\textit{D.~discoideum}) cells interacting with colloidal ``cargo'' beads, similar to concertgoers with large balloons. We have previously studied this system with a different bead size~\cite{Grossmann2024} and found Fickian, yet non-Gaussian diffusion due to heterogeneity in the system. Here, we use a likelihood-based approach to infer the heterogeneity in the dynamics of the cargo beads. Since this method can handle information-scarce situations, we use it to analyze short time snippets and through this uncover a drift of the heterogeneity over the course of the experiment. 

The remainder of the paper is structured as follows. In Sec.~\ref{sec:model}, we describe the system and deduce a model that captures the dynamics and heterogeneity of the beads. In Sec.~\ref{sec:Inference}, we detail the likelihood method and use it for inference on the whole dataset before we investigate the time dependence in Sec.~\ref{sec:non-stationarity}. We end with some more general remarks in Sec.~\ref{sec:conc}.
	\section{Data and model}
\label{sec:model}
We study a composite system of motile ameboid cells and passive polystyrene beads:~On top of a dense monolayer of \textit{D.~discoideum} cells, a small number of beads are added~\cite{Grossmann2024}. Interactions between the cells and the beads are mediated through unspecific adhesion~\cite{Loomis2012} once a cell and a bead come into contact~\cite{lepro_optimal_2022}. The beads experience fluctuating forces through locomotion of the cells as well as their membrane movements~\cite{sharifi_cargo_2024}. Each of the cell-cargo bonds can break either spontaneously~\cite{Nagel2019} or through forces applied by other cells~\cite{sharifi_cargo_2024}. In this way, the layer of cells acts as an active bath for the beads that exhibit non-thermal stochastic trajectories. 

The same experimental setup has been featured in Ref.~\cite{Grossmann2024}. There, a dataset with bead diameter of~$46\, \upmu\mathrm{m}$ was analyzed. Here, we focus on an experiment with beads~$61\, \upmu\mathrm{m}$ in diameter. We perform time-lapse recordings by imaging the system every~$15 \, \mathrm{s}$ during a total recording time of 4 hours. The position of the beads was tracked, leading to a total of~$N = 177$ trajectories. While we only present the~$61\, \upmu\mathrm{m}$ dataset in the main text, the results of our analysis on the dataset from Ref.~\cite{Grossmann2024} as well as two additional datasets with bead diameters of~$29\, \upmu\mathrm{m}$ and~$100\, \upmu\mathrm{m}$ are shown and discussed in the Supplemental Material~\cite{noteSI}. The bead sizes are publisehd in Ref.~\cite{Dautzenberg2025}.

First, we perform a statistical analysis analogous to Ref.~\cite{Grossmann2024}, which will lead to the same conclusions about the form of the model to describe the dynamics of the population of beads. Once the model type is chosen, the following sections focus on the likelihood-based analysis of the data, which is specific to the paper at hand.

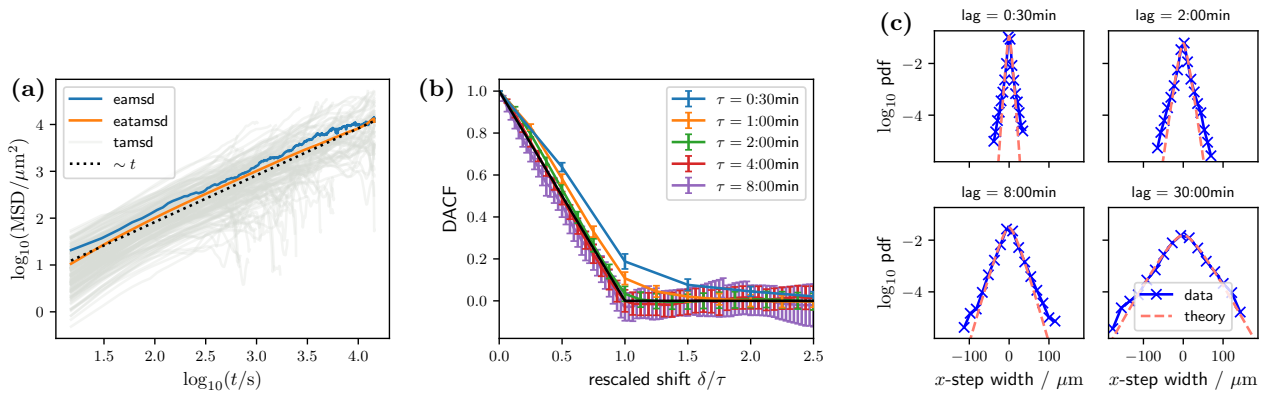
\begin{figure*}[tb]
	\begin{center}
		\includegraphics[width=0.31\textwidth]{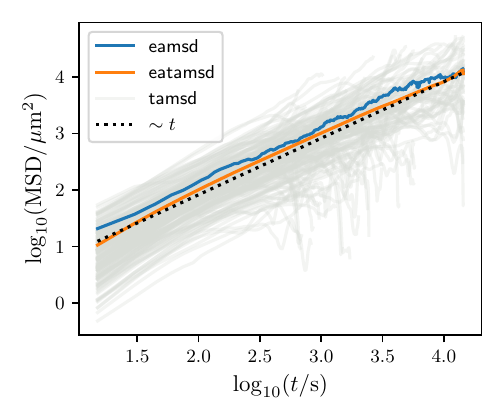}
		\begin{picture}(0,0)
			\put(-0.29\textwidth,125){\makebox(0,0)[lt]{\hspace{-0.6em}\textbf{(a)}}}
		\end{picture}
		\includegraphics[width=0.31\textwidth]{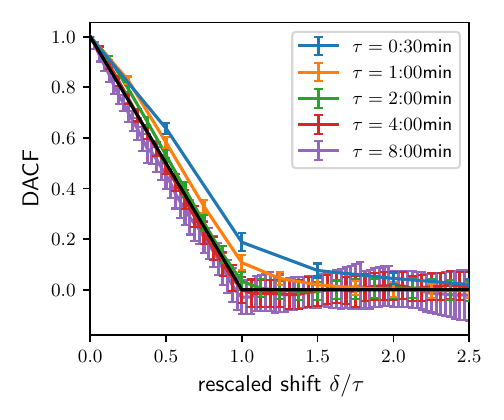}
		\begin{picture}(0,0)
			\put(-0.31\textwidth,125){\makebox(0,0)[lt]{\hspace{-0.6em}\textbf{(b)}}}
		\end{picture}
		\includegraphics[width=0.31\textwidth
		]{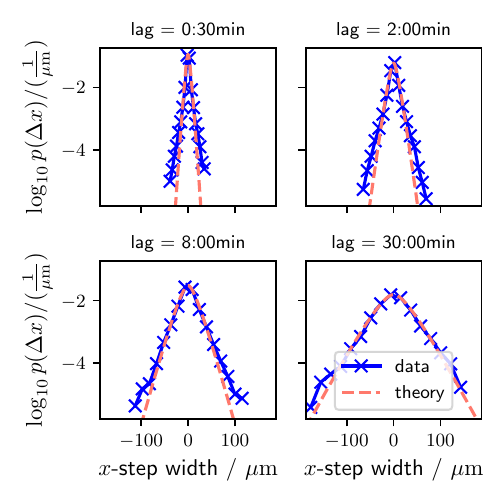}
		\begin{picture}(0,0)
			\put(-0.29\textwidth,155){\makebox(0,0)[lt]{\hspace{-0.6em}\textbf{(c)}}}
		\end{picture}
	\end{center}
        \vspace{-0.6cm}
	\caption{Statistical characterization of bead trajectories driven by an active cell bath. (a) Mean-squared displacement. The blue and the orange lines show \textit{eamsd} and \textit{eatamsd}, respectively. The transparent gray lines show the \textit{tamsd} of the individual trajectories (see Eq.~\eqref{eq:data_tamsd}). The dotted black line shows the theoretical \textit{tamsd} of a long trajectory of a purely diffusive particle, which is proportional to~$t$. (b) Rescaled and normalized displacement autocorrelation functions. The black line indicates the theoretical result for a purely diffusive particle. (c) Step-size distributions in the~$x$-direction. The blue curves are histograms calculated from the dataset. Each cross indicates the center of a bin interval. The red dashed lines are the predictions from the heterogeneity inference~(see Sec.~\ref{sec:Inference}).}
	\label{fig:data_char}
\end{figure*}

We particularly consider the mean-squared displacement~(MSD) of the trajectories. For each trajectory, the squared displacement for a time delay~$t$ with respect to some reference time~$s$ is given by
\begin{gather}
	\omega(t; s) = \left|\vec{r}(t + s) - \vec{r}(s)\right|^2\,.
\end{gather}
The time-averaged mean-squared displacement~(\textit{tamsd}) for a trajectory of duration~$T$ starting at~$t_0=0$ is then given by 
\begin{gather}
	\overline{\omega(t)} = \frac{1}{T-t} \int_0^{T - t}\!\ddint{s} \omega(t;s)\,.
    \label{eq:data_tamsd}
\end{gather}
We also consider the ensemble averages of the squared displacements as well as those of the \textit{tamsd} over all trajectories, which leads to the ensemble-averaged MSD~(\textit{eamsd})~$\expval*{\omega(t)}$ and the ensemble-averaged time-averaged MSD~(\textit{eatamsd})~$\expval*{\overline{\omega(t)}}$, respectively. In case of the \textit{eamsd}, the reference time is the starting time of each trajectory. Figure~\ref{fig:data_char}(a) shows the different types of MSD calculated from the dataset. Starting from~$t\approx 100\,\textrm{s}$, the \textit{eatamsd} has an exponent close to~$1$ (Fickian diffusion) over 2 orders of magnitude, indicating diffusive behavior at time scales longer than 2 min. The \textit{eamsd} is more noisy but appears to be linear in time for even shorter time scales.

In addition to the MSD, we consider the displacement autocorrelation function~(DACF). For a trajectory of duration~$T$ starting at~$t_0=0$ and an interval length of~$\tau$, it is given by
\begin{gather}
	C_\tau(\delta) = \frac{1}{T- (\tau + \delta)}\int_0^{T - (\tau+\delta)}\!\!\! \ddint{t} \Big [ \Delta_\tau \vec{r}(t) \Big ] \! \cdot \! \Big [ \Delta_\tau\vec{r}(t+\delta) \Big] \,,
	\label{eq:data_dacf}
\end{gather}
where~$\Delta_\tau \vec{r}(t) \defeq \vec{r}(t + \tau) - \vec{r}(t)$. We focus on the normalized correlation with rescaled time shift~$\tilde{\delta}\defeq \delta/ \tau$:
\begin{gather}
	\tilde{C}_\tau(\tilde{\delta}) \defeq C_\tau(\tilde{\delta} \cdot \tau) / C_\tau(0)\,.
\end{gather}
The correlation~$\tilde{C}_\tau(\tilde{\delta})$ calculated from the data is shown in Fig.~\ref{fig:data_char}(b) for a number of values for~$\tau$. For interval lengths~$\tau$ larger than some critical value~$\tau_c$, the curves collapse onto one master curve~(see also Ref.~\cite{Grossmann2024}). This master curve is given by the theoretical result of the correlation for a purely diffusive particle. For such a particle, the correlation is proportional to the overlap of the two intervals in Eq.~\eqref{eq:data_dacf} and therefore drops to zero for~$\delta / \tau >1$. The plot indicates that the value of~$\tau_c$ for the experimental bead-cargo system lies between 1 and 2 min. This means that non-overlapping displacements of the beads with interval lengths~$\tau$ above 2:00 min are
independent of each other. This is another indication that the dynamics of the beads is well described by Brownian motion above a time scale of 2:00 min. \changed{Many other models that could have been good candidates \textit{a priori}, like L\'evy walks or flights or processes driven by correlated noise, either contradict one of the above observations or become equivalent to pure diffusion for sufficiently long measurement intervals.}

Finally, we look at the empirical displacement distributions of the dataset at different time lags~$\tau$. The results for steps in the~$x$-direction are shown in Fig.~\ref{fig:data_char}(c); the distributions for~$y$-steps~(not shown) are qualitatively identical. 
In contrast to the results discussed in the paragraphs above, the displacement distributions do not show a signature of Brownian motion: While the displacement distributions of a purely diffusive particle are Gaussian, the experimental step-size distributions show clear exponential tails. 

Going back to the plot of the MSDs in Fig.~\ref{fig:data_char}(a), we observe that the \textit{tamsd} curves are not tightly packed as would be expected for a homogeneous ensemble but are rather spread out over almost 2 orders of magnitude. This spread hints at heterogeneity in the system. Variability in the dynamics of the individual beads can arise through random variations in the local cell density as well as through variability of the dynamics of the cells that drive the beads. Let us assume that the dynamics of each of the beads above a time scale of 2:00 min is well captured by standard diffusion
\begin{gather}
	\dot{\vec{r}}^n(t) = \sigma_n \, \vec{\xi}^n(t)\,,
	\label{eq:data_diffusion}
\end{gather}
as suggested by the MSDs and the DACF. Here, the index~$n\in[1,\ldots, N]$ indicates the individual beads and~$\vec{\xi}^n(t)$ are independent white noise processes. We can then estimate the noise strength~$\sigma_n^2$ for each of the trajectories using the maximum likelihood estimator given in Eq.~\eqref{eq:AppThy_noisestrengthMLE}. The empirical distribution of these noise strengths shown in Fig.~\ref{fig:inference_results}(b) has a single peak and is well approximated by the probability density function~(pdf) of a~$\Gamma$-distribution. We therefore model the system as diffusive particles driven by the stochastic differential equation~\eqref{eq:data_diffusion} with noise strengths~$\left\{\sigma^2_n\right\}_{n=1,\ldots,N}$ identically and independently distributed according to a~$\Gamma$-distribution 
\begin{gather}
	\sigma^2_n \sim \Gamma(\alpha, \beta)\,,
	\label{eq:model_heterogeneity}
\end{gather}
with pdf
\begin{gather}
	q_{\Gamma}(x|\alpha, \beta) = \frac{\beta^\alpha}{\Gamma(\alpha)}x^{\alpha -1}e^{-\beta x}\,	
\end{gather}
and heterogeneity parameters~$\boldsymbol{\theta} \defeq (\alpha, \beta)$.
While other distributions might also fit the heterogeneity of the noise strengths, 
the use of the~$\Gamma$-function is motivated by its analytical properties, which ease the calculations in the following sections. \changed{Other choices do not qualitatively change the conclusions presented in the following \cite{noteSI}.}

In the next section, we will use the model comprised of Eqs.~\eqref{eq:data_diffusion} and \eqref{eq:model_heterogeneity} to estimate the heterogeneity of the experimental system.


	\section{Likelihood-based inference}
\label{sec:Inference}

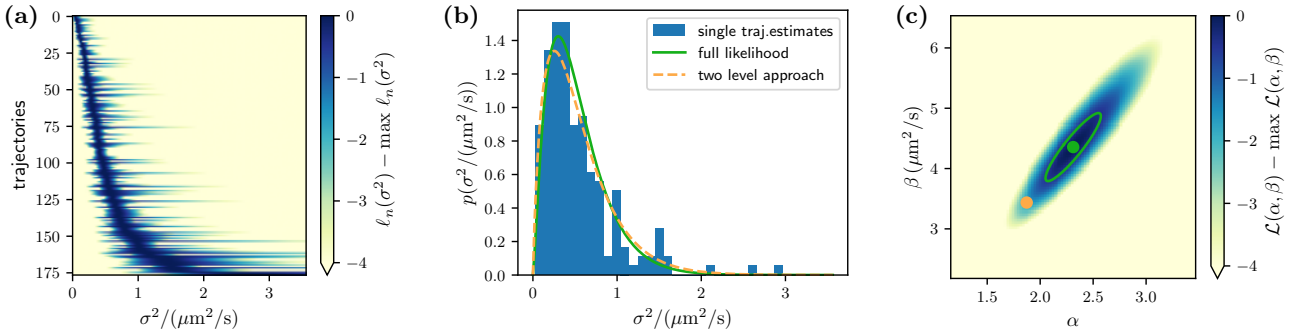
\begin{figure*}[tb]
	\centering
	\includegraphics[height=0.55\columnwidth]{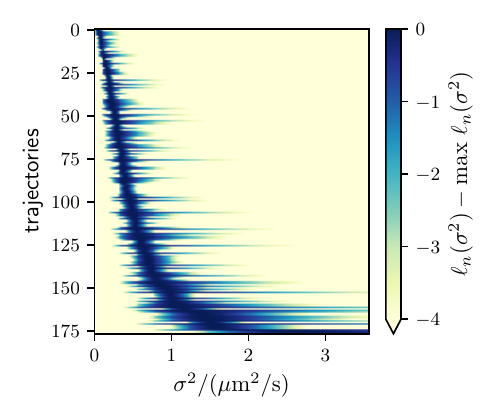}
	\begin{picture}(0,0)
		\put(-0.305\textwidth,130){\makebox(0,0)[lt]{\hspace{-0.6em}\textbf{(a)}}}
	\end{picture}
	\includegraphics[height=0.55\columnwidth]{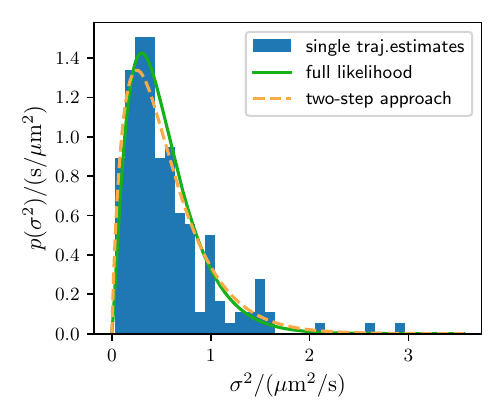}
	\begin{picture}(0,0)
		\put(-0.31\textwidth,130){\makebox(0,0)[lt]{\hspace{-0.6em}\textbf{(b)}}}
	\end{picture}
	\includegraphics[height=0.55\columnwidth]{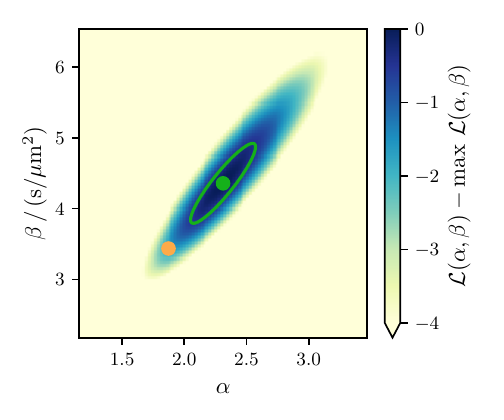}
	\begin{picture}(0,0)
		\put(-0.305\textwidth,130){\makebox(0,0)[lt]{\hspace{-0.6em}\textbf{(c)}}}
	\end{picture}
	\vspace{-0.3cm}
	\caption{Heterogeneity inference on bead trajectories driven by an active cell bath. A sparse version of each trajectory with~$\Delta t= \text{2~min}$ is considered to ensure that the dynamics are in the diffusive regime. (a) Log-likelihood of the noise strength given the trajectories of the beads (see Eq.~\eqref{eq:Inf_noisestrengthllh}). The trajectories are sorted by their maximum likelihood estimate for the noise strength. (b) Inferred probability density function of the noise strengths. The green curve represents the result of the full likelihood approach, while the dashed orange curve is the result from a two-step approach. The histogram shows the MLEs of the noise strengths based on the functions shown in panel~(a). (c) Log-likelihood~$\mathcal{L}(\vec{\theta})$ of the heterogeneity parameters with respect to the sparse dataset. The green dot denotes the MLE~$\hat{\vec{\theta}}$ that maximizes~$\mathcal{L}(\vec{\theta})$. The green ellipse is the uncertainty estimate calculated from the Hessian matrix at~$\vec{\theta}$. The orange dot denotes the estimate obtained from a two-step inference approach.} 
	\label{fig:inference_results}
\end{figure*}

In the previous section, we have derived a hierarchical model: On the lower level, the noise strength~$\sigma^2_n$ for each of the individual trajectories is unknown and needs to be inferred from the position data; on the higher level, the system heterogeneity, i.e., the statistical distribution of these unknown noise strengths over the dataset, is also unknown. 

There are at least two distinct approaches to estimate the parameters of the heterogeneity distribution~\cite{Albrecht2024}. One is to treat the two levels of stochasticity individually. This means that, first, estimators~$\hat{\sigma}_n^2$ are calculated independently for each trajectory. In a second step, these estimates can then be used within some inference scheme to estimate the heterogeneity. We call this approach, which has been used in Ref.~\cite{Grossmann2024}, a two-step estimate. Such a procedure is also known as a two-stage~\cite{GenonCatalot2016} or plug-in~\cite{Delattre2015} approach. When doing so, all information included in a trajectory is reduced to a single number~$\hat{\sigma}_n^2$ and any uncertainty about this estimate is ignored. One way to quantify the range of plausible values for the noise strength given a trajectory~$\vec{T}^n$ is to look at the log-likelihoods of the noise strengthÖ
\begin{gather}
	\log p(\vec{T}^n|\sigma_n^2) \eqdef \ell_n(\sigma_n^2)\,.
	\label{eq:Inf_noisestrengthllh}
\end{gather}
For Brownian motion as described by Eq.~\eqref{eq:data_diffusion}, it can be expressed as a product of Gaussians~\cite{Risken1996}; the full expression of~$p(\vec{T}^n|\sigma_n^2)$ is given in Eq.~\eqref{eq:AppThy_noisestrengthllh}.
In Fig.~\ref{fig:inference_results}(a), the log-likelihoods~$\ell_n(\sigma_n^2)$ with respect to the trajectories in the experimental dataset are shown. Some of the log-likelihood functions are narrow with respect to the range of noise strengths in the dataset. For these trajectories, the reduction of the information in the trajectory to the one estimate value~$\hat{\sigma}_n^2$ is justified. However, for other trajectories,  the likelihood is very broad, indicating that there is a wide range of possible and relevant values for the noise strength.

In order to use the available information efficiently, an inference scheme for the heterogeneity should take this range of possible values into account. We therefore propose here to use a full likelihood approach that allows inference of the heterogeneity parameters directly from the measured positions~\cite{Albrecht2024}. While Eq.~\eqref{eq:Inf_noisestrengthllh} is the log-likelihood of the noise strength with respect to a trajectory, we now consider the log-likelihood of the heterogeneity parameters~$\vec{\theta}=(\alpha, \beta)$: 
\begin{gather}
	\log 	p(\vec{T}^n|\vec{\theta}) = \log \!\int \ddint{\sigma^2} p(\vec{T}^n|\sigma^2) \,q_{\Gamma}(\sigma^2|\vec{\theta})\eqdef \mathcal{L}_n(\vec{\theta})\,.
	\label{eq:Inf_trajlikelihood}
\end{gather}
The integration inside the logarithm combines all possible noise intensities~$\sigma^2_n$, each weighted by the likelihood. This way, an information bottleneck as for the two-step approach is avoided. 

The likelihoods for the individual trajectories can be combined into a log-likelihood of~$\vec{\theta}$ with respect to the full dataset $\mathcal{D} = \left\{\vec{T}^n\right\}_{n=1, \ldots, N}$:
\begin{gather}
	\mathcal{L}(\vec{\theta})\defeq\log p(\mathcal{D}|\vec{\theta}) = \sum_n\mathcal{L}_n(\vec{\theta})\,.
	\label{eq:Inf_likelihood}
\end{gather}
The parameter value that maximizes a likelihood function~$\hat{\vec{\theta}}$ is known as the maximum likelihood estimator. This estimator has favorable statistical properties like consistency and
efficiency \cite{Schervish1995}. Furthermore, just like the shape of the noise strength log-likelihood was indicative of the range of relevant noise strengths, the Hessian matrix of~$\mathcal{L}(\vec{\theta})$ at the MLE is connected to the uncertainty of~$\hat{\vec{\theta}}$: For a large number of trajectories, the negative inverse of the Hessian matrix approaches the variance matrix of the MLE~\cite{Schervish1995, Albrecht2024}. 
For the model at hand, the integral in Eq.~\eqref{eq:Inf_trajlikelihood} and the elements of the Hessian matrix have analytical expressions in terms of modified Bessel functions and their derivatives (see Appendix \ref{sec:AppTheory} and Ref.~\cite{noteSI}). There is no analytic expression for~$\hat{\vec{\theta}}$, but it can be easily obtained by numerical maximization of~$\mathcal{L}(\vec{\theta})$.

We apply this full likelihood approach to the dataset of bead trajectories in the experimental cell-cargo system. As described in Sec.~\ref{sec:model}, the system has been imaged every~$15\,\mathrm{s}$, but the observed dynamics can only be assumed to be diffusive above a time scale of 2 min. We therefore work with sparse versions of the trajectories in which positions are removed such that the time difference between subsequent positions in the modified trajectory is 2 min. When creating these sparse trajectories, we start from the first position in every trajectory. In the following, all inference results will be based on this sparse dataset. 

\begin{figure*}[htb]
	\includegraphics[height=0.55\columnwidth]{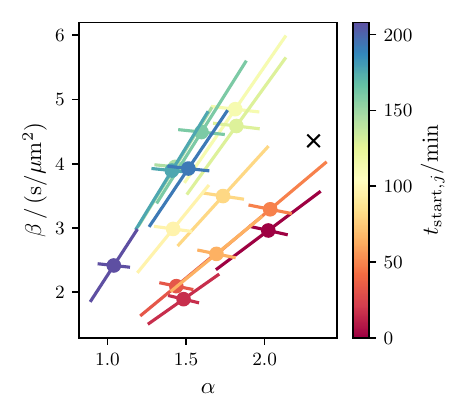}
	\begin{picture}(0,0)
		\put(-0.285\textwidth,130){\makebox(0,0)[lt]{\hspace{-0.6em}\textbf{(a)}}}
	\end{picture}
	\includegraphics[height=0.55\columnwidth]{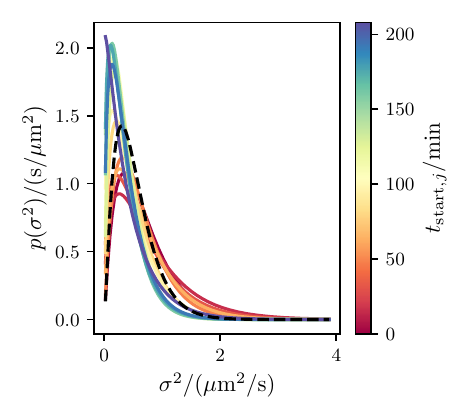}
	\begin{picture}(0,0)
		\put(-0.285\textwidth,130){\makebox(0,0)[lt]{\hspace{-0.6em}\textbf{(b)}}}
	\end{picture}
	\includegraphics[height=0.55\columnwidth]{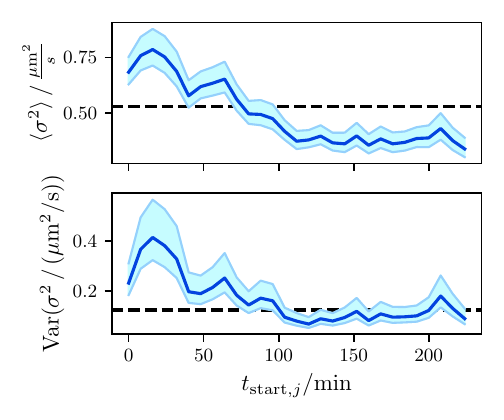}
	\begin{picture}(0,0)
		\put(-0.31\textwidth,135){\makebox(0,0)[lt]{\hspace{-0.6em}\textbf{(c)}}}
	\end{picture}
	\vspace{-0.3cm}
	\caption{Time dependence of heterogeneity inference. Inference was performed on reduced datasets containing only data points within a time window of 16~min. The start time~$t_{\text{start},j}$ of this time window was shifted over the observation time of the experiment. (a) Inferred heterogeneity parameters~$\hat{\vec{\theta}}_j$. The black cross denotes the result of the inference from the complete dataset. Note that the error bars appear non perpendicular due to the skewed aspect ratio. (b) Inferred heterogeneity distribution. The black dashed line denotes the result of the inference from the complete dataset, reflecting an average behavior. (c) Mean and variance of the inferred distribution over time. The dashed line denotes the result of the inference from the complete dataset. The plotted error bounds correspond to the highest and lowest values of mean and variance within the 1$\sigma$ uncertainty bounds in $\vec{\theta}$ space.  In panel~(c), the plotted~$t_{\text{start},j}$ are spaced 8~min apart, which means that the windows partially overlap. Note that in panels~(a) and (b) only every second point is plotted for better readability, leading to a spacing of 16~min.}
	\label{fig:fluc_results}
\end{figure*}
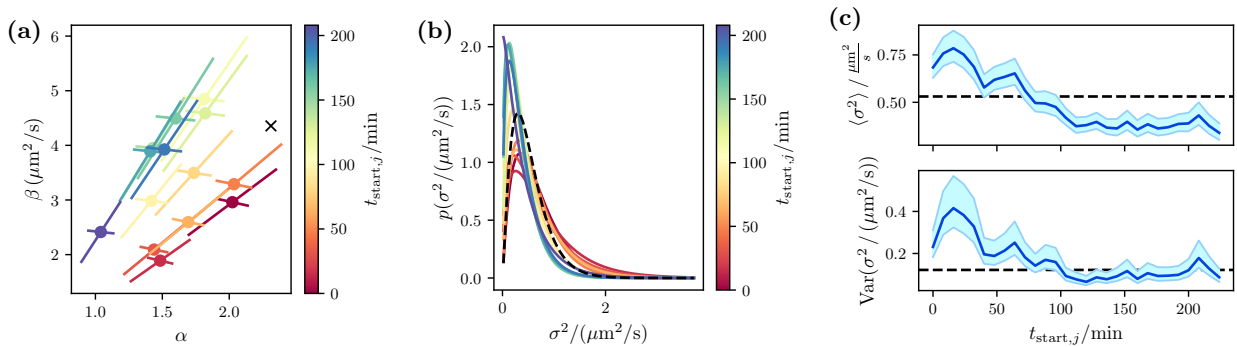

Figure \ref{fig:inference_results}(c) shows the log-likelihood of the heterogeneity parameters with respect to the sparse experimental data. Note that~$\alpha$ is a dimensionless number, while~$\beta$ has inverse units of the noise strength. In addition, the MLE~$\hat{\vec{\theta}}$ is displayed as well as an ellipse that denotes the uncertainty estimate. The corresponding density of the heterogeneity distribution is shown in Fig.~\ref{fig:inference_results}(b). The inference confirms that the effective noise strengths that drive the beads are strongly heterogeneous over the dataset. Similarly to Ref.~\cite{Grossmann2024}, we confirm that the inferred heterogeneity explains the behavior of the step size distributions in Fig.~\ref{fig:data_char}, except for the outmost tails. The orange dotted line and the orange dot in Figs.~\ref{fig:inference_results}(b) and (c), respectively, show an alternative heterogeneity estimate from the two-step approach. After calculating estimates of the noise strength for each trajectory by maximizing Eq.~\eqref{eq:Inf_noisestrengthllh}, these estimates~$\left\{\hat{\sigma}_n^2\right\}$ are used in a maximum likelihood scheme to optimize a~$\Gamma$-distribution. The estimator from the two-step approach is outside the uncertainty bounds of the full-likelihood approach. 

The need for reliable uncertainty estimates and the likelihood framework increases when less data are available, since less data imply a larger uncertainty about the individual noise strength estimates. In the next section, we will apply the approach to a situation in which the available data are limited, namely, short trajectories:~We analyze only snippets of trajectories to assess the stationarity of the bead dynamics. 

	\section{Non-stationarity of the system}
\label{sec:non-stationarity}
The model in Eqs.~\eqref{eq:data_diffusion} and \eqref{eq:model_heterogeneity} for the dynamics of the beads and their heterogeneity that we have used for inference in Sec.~\ref{sec:Inference} is stationary---The noise strengths and therefore also the heterogeneity parameters are assumed not to change over time. In the following, we will relax this assumption. 

While the results presented in Fig.~\ref{fig:data_char} strongly suggest diffusive dynamics of the beads, it is still possible that the value of the noise strength changes on timescales longer than the measurement frequency~\cite{Chubynsky2014}. Over the duration of the experiments, the beads are transported from one part of the cell carpet to the next, where other distinct cells are attached to the bead and also the cell density might be different. Furthermore, the behavior of the cells might change over time. Here, we are not primarily interested in the exact dynamics of the individual beads, but rather want to know if the overall heterogeneity distribution of the dynamics in the system changes over the course of the experiment, i.e., whether the system shows non-stationary behavior. 

In order to do so, we consider short time windows with onset time~$t_{\text{start}, j}$ and duration~$w=\text{16~min}$ within which we assume the system to be stationary. 
%
The motion of individual \textit{D.~discoideum} cells is diffusive above a timescale of around $10\,\mathrm{min}$ with an average diffusion coefficient lower than $10\upmu\mathrm{m}^2/\mathrm{min}$ \cite{Grossmann2024, Gole2011}. 
This means their expected displacement in $16\,\mathrm{min}$ is $(4\cdot 10\upmu\mathrm{m}^2/\mathrm{min} \cdot16\mathrm{min})^{1/2}\approx 25\upmu\mathrm{m}$ while being only~$10$--$20\,\upmu\mathrm{m}$ in diameter~\cite{Annesley2009}.
Therefore, only limited rearrangement of the active bath is possible during these time windows.
The heterogeneity can then be estimated using the data in this time window only. Note that there will be a maximum of only nine positions per trajectory within each window width since we are using sparse trajectories with 2~min between data points. For each window~$j$, the full likelihood approach returns an estimate for the heterogeneity parameters~$\hat{\vec{\theta}}_j$ together with an uncertainty estimate. By shifting the time window, we obtain a series of estimates~$\{\hat{\vec{\theta}}_j\}$ and their uncertainties.
Note that the full likelihood approach is particularly well suited to deal with these short trajectories, since heterogeneity distributions can be inferred taking into account fluctuations at the single trajectory level under these information-scarce conditions, whereas a two-step approach (see Appendix \ref{sec:App2Level}) is prone to inaccuracy as detailed in Sec.~\ref{sec:Inference}.

The results in Fig.~\ref{fig:fluc_results} show that the system is indeed not stationary but transient. The mean and the variance of the heterogeneity distribution decrease over the first two hours of the experiment before becoming much more stable during the second two hours. 
The dynamics in~$\vec{\theta}$ space with their uncertainties depicted in Fig.~\ref{fig:fluc_results}(a) show that the change in the first half of the experiment is indeed a significant, systematic shift rather than random fluctuations. During the second half, there is a large overlap of the uncertainties of the estimates. This indicates that the system has reached a quasi-stationary state.
In Fig.~\ref{fig:fluc_results}(b), we can see that the heterogeneity distribution estimated from the full dataset constitutes a midpoint between the distributions inferred at the beginning and the end of the experiment. \changed{The results above are robust against the changes of the window duration \cite{noteSI}.}
 
The drop in the average of the distribution means that the diffusion coefficient of the beads decreases over time. One important factor is probably the accumulation of cells around the beads, which is also visible in the raw footage from the microscope~\cite{noteSI}. The accumulation arises due to the unspecific adhesion of \textit{D.~discoideum} cells~\cite{Loomis2012} to the beads and, moreover, cells prefer structured environments with more contact area over flat surfaces~\cite{Arcizet2012}. Such an accumulation has several effects. 
The more cells are attached to a bead, the more the forces, which cells exert on the bead, cancel each other. 
Furthermore, it is less likely for a single cell to pull the bead off all other cells and move it by a larger distance. Both effects decrease the effective diffusion constant of the bead. Another aspect is that the maximum number of cells around a bead is limited by the steric repulsion of cells. Thus, the accumulation will naturally saturate at some density, which is similar for all beads. This decreases the contribution of cell density fluctuations to the variability of the dynamics of the beads, explaining the reduced variance. 

Another possible contributing factor to the temporal change of the heterogeneity is a time dependence of the behavior of the cells themselves. \textit{D.~discoideum} cells are known to show quorum-sensing behavior that affects their motility~\cite{Gole2011}. The increase in concentration of the quorum-sensing factor could lead to a slowdown in their dynamics until a saturation of the sensing pathway sets in. A generally slower movement of each cell would lead to decreased mean as well as a decreased variance of the bead dynamics. Tracking and analyzing the motility of the cells would give a clearer picture about the influence of this factor; however, this is beyond the scope of the present study. 



The non-stationarity of the data does not come as a surprise when we take a closer look at the MSD curves in Fig.~\ref{fig:data_char}(a). For a stationary system, the additional time-averaging in the \textit{eatamsd} leads to a less noisy curve compared to the \textit{eamsd} but does not shift the curve. The discrepancy of \textit{eamsd} and \textit{eatamsd} therefore indicates changes of the dynamics over time. The thorough likelihood-based analysis of the time-windows reveals the nature of the non-stationarity in the system as well as the quasi-stationary behavior in the second half of the experiment. 

	\section{Conclusion}
\label{sec:conc}
We have presented a likelihood-based approach to study the time-dependent heterogeneous dynamics of micro-beads in a bath of active ameboid cells. The likelihood approach allows to skip the intermediate estimation of trajectory-specific quantities---the individual effective diffusion coefficients in this case. This way, the available trajectory information is used efficiently and information bottlenecks are avoided. In addition, the approach provides natural uncertainty estimates that allow to judge the reliability of the heterogeneity estimates.

Using the likelihood-based approach, we show that the heterogeneity within the motion pattern of the population of beads changes significantly over the course of the experiment. During the first two hours, the dynamics ``cool off,'' becoming slower and less variable before reaching a quasi-stationary state for the remainder of the experiment. When conducting the experiment, the cells were given enough time to attach to the glass surface and start their basic motion pattern. Similarly, it was ensured that the beads had time to sediment before the start of the recording (see Ref.~\cite{noteSI}). Despite the two individual subsystems being equilibrated prior to the recording, the composite system of beads and cells needs much longer to reach a steady state. 

Not accounting for the dynamics of the heterogeneity can skew results when trying to quantify the variability in the system. 
But even when the heterogeneity quantification is not the goal, analyzing its time dependency can still be beneficial. It may be used to identify windows of stationary dynamics which might be challenging to do by analyzing the imaging data or considering the extracted trajectories individually. The likelihood-approach is especially well suited since it can deal with the short trajectories within the time windows.

The system studied here was described by a model that has an analytically tractable likelihood for the heterogeneity parameters. However, the same likelihood machinery can also be used on more complex systems with more involved models for which the exact likelihoods are untractable. In such cases, approximation techniques are required~\cite{Albrecht2024, DelattreLavielle2013}, but once appropriate likelihood expressions have been found the method itself is also able to deal with additional terms such as particle-particle~\cite{Ariel2015} or particle-wall~\cite{Lambert2025} interactions. The more complex the model, the more difficult it is to estimate parameters and variability from limited data. Especially here, the likelihood approach can help to uncover heterogeneities as well as time-dependence of the parameters.

	\section*{Acknowledgments}
We thank Kirsten Sachse for supporting laboratory
routines.
This research has been partially funded by the Deut\-sche For\-schungs\-ge\-mein\-schaft~(DFG) -- Project-ID 318763901 -- SFB1294.
\section*{Data Availability}
The data that support the findings of this article are openly available \cite{Dautzenberg2025}.
	\appendix
	\section{Theoretical Modeling}
\label{sec:AppTheory}
We consider an ensemble of~$N$ Brownian particles. The dynamics of each of them are described by 
\begin{gather}
	\dot{\vec{r}}^n(t) = \sigma_n \, \vec{\xi}^n(t)\,,
\end{gather}
where~$\vec{\xi}^n(t)$ is~$d$-dimensional white noise and~$\sigma_n^2$ is the particle-specific noise intensity. These noise strengths are identically and independently distributed according to a~$\Gamma$-distribution~$\Gamma(\alpha, \beta)$ with density
\begin{gather}
	q_{\Gamma}(\sigma^2) = \frac{\beta^\alpha}{\Gamma(\alpha)}\, e^{-\sigma^2 \beta}\, \left(\sigma^2\right)^{\alpha - 1}\,.
\end{gather}
The position of the particles is measured discretely at certain points in time, leading to observed trajectories of the form
\begin{gather}
	\vec{T}^n = \left\{(t^n_i, \vec{r}^n_i)\right\}_{i = 1, \ldots, M^n + 1}\,.
\end{gather}
Note that the upper index~$n$ indicates the particle, while the lower index~$i$ labels the specific measurement. 
This leads to observed displacements~$\Delta \vec{r}^n_i \defeq \vec{r}^n_{i+1} - \vec{r}^n_i$ and corresponding time lags~$\Delta t^n_i \defeq t^n_{i+1} - t_i^n$. A trajectory contains~$M^n +1$ positions and therefore~$M^n$ displacements. 

Given the noise strength~$\sigma_n^2$, the probability of a measured trajectory follows from the Brownian dynamics and is
\begin{gather}
	\! p(\vec{T}^n|\sigma_n^2) = \prod_{i=1}^{M^n}\left(2\pi\, \sigma_n^2 \Delta t^n_i\right)^{-\frac{d}{2}} \exp\!\left(- \frac{\left(\Delta\vec{r}^n_i\right)^2}{2\,\sigma_n^2\,\Delta t^n_i}\right)\! , \! 
	\label{eq:AppThy_noisestrengthllh}
\end{gather}
where~$d$ is the spatial dimension of the process. For the experimental system considered in this paper, we have~$d=2$. Given a trajectory~$\vec{T}^n$, Eq.~\eqref{eq:AppThy_noisestrengthllh} is maximized by
\begin{gather}
	\hat{\sigma}_n^2 \defeq \frac{1}{d\,M^n}\sum_i\frac{\left(\Delta\vec{r}^n_i\right)^2}{\Delta t^n_i}\,,
	\label{eq:AppThy_noisestrengthMLE}
\end{gather}
which is the maximum likelihood estimator (MLE) for the noise strength.

The probability of a trajectory conditioned on the parameters of the~$\Gamma$-distribution is calculated by
\begin{gather}
	p(\vec{T}^n|\alpha, \beta) = \int \ddint{\sigma^2} p(\vec{T}^n|\sigma^2) \,q_{\Gamma}(\sigma^2|\alpha, \beta)\,.
\end{gather}
This is just the likelihood of the heterogeneity parameters~$(\alpha, \beta)$ given trajectory~$\vec{T}^n$.
The above integral has an explicit analytical solution, which is given by
\begin{align}
	\log p(\vec{T}^n|\alpha, \beta) = & C +  \frac{1}{2}\left(\alpha - b_n\right) \log \left(\frac{a_n}{\beta}\right) + \alpha\log\beta \nonumber\\
	&- \log \Gamma(\alpha) + \log K_{(b_n -\alpha)}(2\sqrt{a_n\, \beta}) \nonumber\\
	\eqdef& \mathcal{L}_n(\alpha, \beta)\,,
	\label{eq:Theo_singleTrajL}
\end{align}
with shorthands
\begin{subequations}
	\begin{align}
		C \defeq& \log 2 - b_n \log\,2 \pi - \frac{d}{2}\sum_{i=1}^{M^n}\log \Delta t_i^n\,,\\
		a_n \defeq &\frac{1}{2}\sum_{i=1}^{M^n} \frac{\left(\Delta\vec{r}_{i}^n\right)^2}{\Delta t_i^n}\,,\\
		b_n \defeq &\frac{M^n\, d}{2}\,,
	\end{align}
\end{subequations}
and where~$K_\nu(x)$ is the modified Bessel function of the second kind. 
The likelihood of the heterogeneity parameters with respect to the whole dataset $\mathcal{D} = \left\{\vec{T}^n\right\}_{n=1,\ldots,N}$ is then
\begin{gather}
	\mathcal{L}(\alpha, \beta) = \log p(\mathcal{D}| \alpha, \beta) = \sum_{n=1}^N \mathcal{L}_n(\alpha, \beta)\,.
	\label{eq:model_llh}
\end{gather}
The maximum of Eq.~\eqref{eq:model_llh} is the maximum likelihood estimate~$(\hat{\alpha}, \hat{\beta})$ for the heterogeneity parameters~$\alpha$ and~$\beta$. 

In order to estimate the uncertainty of this estimate, we calculate the Hessian matrix of~$\mathcal{L}(\alpha, \beta)$ at~$(\hat{\alpha}, \hat{\beta})$. This can be understood as an empirical approximation of the negative Fisher information matrix~$-I_{ij}(\hat{\vec{\theta}}) = \left.\mathbb{E}[\partial_{\theta_i}\partial_{\theta_j}\mathcal{L}(\vec{\theta})]\right|_{\vec{\theta}=\hat{\vec{\theta}}}$, where the expectation value is with respect to the true distribution of the data. \changed{In the limit of many trajectories per dataset, the maximum likelihood estimates for different realizations of a dataset are expected to be normally distributed around the true value with the variance matrix being the inverse Fisher information matrix $I^{-1}(\hat{\vec{\theta}})$ \cite{Schervish1995, Albrecht2024}. In two dimensions, around $39.3\%$ of the probability mass lies within the 1$\sigma$ ellipse and $86.5\%$ within the 2$\sigma$ ellipse (see Ref.~\cite{noteSI}).}

The Hessian matrix of~$\mathcal{L}(\alpha, \beta)$ is 
\begin{align}
	H_{\mathcal{L}}(\hat{\alpha}, \hat{\beta}) =& \sum_n H_{\mathcal{L}_n}(\hat{\alpha}, \hat{\beta}) \nonumber\\
	=& \sum_n
	\left.\begin{pmatrix}
		\partial_\alpha\,\partial_\alpha\, \mathcal{L}_n & \partial_\alpha\,\partial_\beta\, \mathcal{L}_n \\
		\partial_\beta\,\partial_\alpha\, \mathcal{L}_n& \partial_\beta\,\partial_\beta\, \mathcal{L}_n
	\end{pmatrix}\right|_{(\alpha, \beta) = (\hat{\alpha}, \hat{\beta}) }\,.
\end{align}
This matrix can be calculated directly by differentiating Eq.~\eqref{eq:Theo_singleTrajL}. In Ref.~\cite{noteSI} we sketch a way that eases implementation.
    \section{Time Dependence Inferred Using a Two-Step Approach}
\label{sec:App2Level}
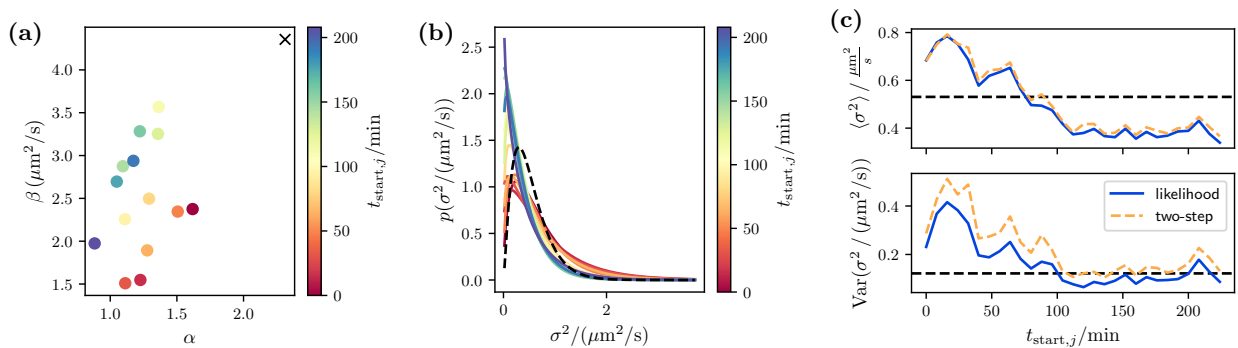
\begin{figure*}
	\includegraphics[height=0.55\columnwidth]{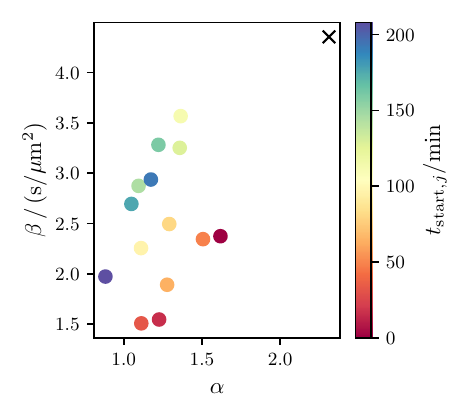}
	\begin{picture}(0,0)
		\put(-0.285\textwidth,130){\makebox(0,0)[lt]{\hspace{-0.6em}\textbf{(a)}}}
	\end{picture}
	\includegraphics[height=0.55\columnwidth]{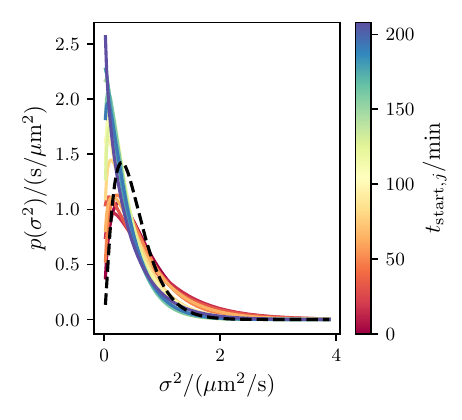}
	\begin{picture}(0,0)
		\put(-0.285\textwidth,130){\makebox(0,0)[lt]{\hspace{-0.6em}\textbf{(b)}}}
	\end{picture}
	\includegraphics[height=0.55\columnwidth]{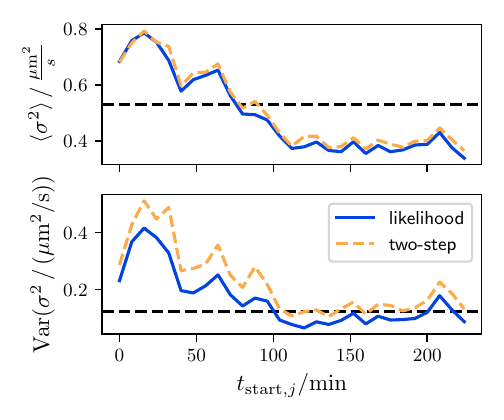}
	\begin{picture}(0,0)
		\put(-0.31\textwidth,135){\makebox(0,0)[lt]{\hspace{-0.6em}\textbf{(c)}}}
	\end{picture}
        \vspace{-0.3cm}
	\caption{Time dependence of heterogeneity inference using a two-step approach with~$\hat{\sigma}^2_n$ obtained from single trajectory MLEs. Inference was performed on reduced datasets containing only data points within a time window of 16~min. The start time~$t_{\text{start},j}$ of this time window was shifted over the observation time of the experiment. (a) Inferred heterogeneity parameters~$\hat{\vec{\theta}}_j$. (b) Inferred heterogeneity distribution. As a reference, the black cross and the black dashed line denote the result of the likelihood inference from the complete dataset. (c) Mean and variance of the inferred distribution over time. The dashed line denotes the result of the likelihood inference from the complete dataset.}
	\label{fig:fluc_results_2step}
\end{figure*}

In Fig.~\ref{fig:fluc_results_2step}, the results of heterogeneity inference on 16:00~min time windows using the two-step approach are shown. The estimates for~$\hat{\sigma}^2_n$ were obtained using the MLE given in Eq.~\eqref{eq:AppThy_noisestrengthMLE}. The heterogeneity estimates were then calculated as MLEs of a~$\Gamma$-distribution with respect to the set of noise strength estimates~$\left\{\hat{\sigma}^2_n\right\}_{n=1,\ldots, N}$. In Fig.~\ref{fig:fluc_results_2step}(c) we can see that the variance clearly deviates, while the estimated mean of the distributions is close to the results of the likelihood inference. Accordingly, we can also observe a shift of the estimates in heterogeneity parameters space displayed in panel (a). Importantly, the heterogeneity estimates have no uncertainty bars associated with them in contrast to the full likelihood-based method.
	
%

\end{document}